\documentclass[
 aip,
 jcp,
 amsmath,
 amssymb,
]{revtex4-1}
\usepackage[utf8]{inputenc}
\usepackage{array}
\usepackage{amssymb}
\usepackage{amsmath}
\usepackage{float}
\usepackage{xcolor}
\usepackage{graphicx}
\usepackage{braket}
\usepackage[hidelinks]{hyperref}

\begin{document}

\title{Orbital-optimized spin-adapted multistate contracted VQE for excited states and properties on quantum hardware}
\author{Erik Rosendahl Kjellgren}
\email{kjellgren@sdu.dk}
\affiliation{Department of Physics, Chemistry and Pharmacy,
University of Southern Denmark, Campusvej 55, 5230 Odense, Denmark.}
\author{Karl Michael Ziems}
\email{k.m.ziems@soton.ac.uk}
\affiliation{School of Chemistry, University of Southampton, Highfield, Southampton SO17 1BJ, United Kingdom}
\affiliation{Department of Chemistry, Technical University of Denmark, Kemitorvet Building 207, DK-2800 Kongens Lyngby, Denmark.}
\author{Peter Reinholdt}
\affiliation{Department of Physics, Chemistry and Pharmacy,
University of Southern Denmark, Campusvej 55, 5230 Odense, Denmark.}
\author{Stephan P. A. Sauer}
\affiliation{Department of Chemistry, University of Copenhagen, DK-2100 Copenhagen \O, Denmark.}
\author{Sonia Coriani}
\affiliation{Department of Chemistry, Technical University of Denmark, Kemitorvet Building 207, DK-2800 Kongens Lyngby, Denmark.}
\author{Jacob Kongsted}
\affiliation{Department of Physics, Chemistry and Pharmacy,
University of Southern Denmark, Campusvej 55, 5230 Odense, Denmark.}
\date{\today}

\begin{abstract}
We introduce the orbital-optimized multistate contracted variational quantum eigensolver (oo-MC-VQE) method with spin-adapted operators for the computation of ground and excited states, as well as state-specific and transition properties. The use of spin-adapted operators ensures that the spin symmetry of the reference states is conserved throughout the VQE optimization.
In multistate variational approaches, achieving a balanced description of an increasing number of electronic states places growing demands on the expressibility of the underlying ansatz, thereby introducing a fundamental trade-off between accuracy and circuit complexity.
We consider the effects of this trade-off explicitly and find that the number of circuit parameters required to obtain accurate results is reported to scale approximately linearly in the number of states.
We further present an explicit quantum-circuit implementation of the oo-MC-VQE method and demonstrate its integration with quantum error mitigation techniques. Finally, we execute the method on real quantum devices to compute absorption spectra for two benchmark molecular systems.
\end{abstract}


\maketitle

\section{Introduction}
The prediction of electronically excited states and their associated transition properties is central to the theoretical description of molecular spectroscopy.\cite{Mustroph2015} Excitation energies and transition moments can be connected to experimentally observable quantities such as absorption spectra and play a key role in understanding photochemical processes, energy transfer, and excited-state dynamics. 
Electronically excited states and their properties are also key to, for example, the design of photoactive organic materials used for applications such as photovoltaics, photocatalysts for solar fuels, light-emitting diodes, wastewater purification, and photodynamic therapy.\cite{Hamblin2004,Cardone2022,Mohamadpour2024,TAO2025102142}
The efficiency of such materials directly depends on the transition rates between different electronic states, and excited state engineering is a direct strategy to improve their efficiency.\cite{Pham2021}

Despite their importance, the robust and balanced computation of these quantities remains a significant challenge, especially in situations involving strong electronic reorganization, near-degeneracies, or multiconfigurational character.

Broadly speaking, two distinct routes can be taken to describe electronically excited states. The first is based on response theory, where excitation energies and transition properties are obtained through linear-response\cite{Olsen1985-pg,Christiansen1998-mb,Pawlowski2015-vd} (LR) formalism built on top of a ground-state reference.\cite{helgaker2012}
This approach has been highly successful in many regimes, but they fundamentally treat excited states as perturbations of the reference state.
As a result, they may encounter conceptual and practical limitations in certain regimes, particularly when the excited states involve strong electronic reorganization or near-degeneracies.
The second route is based on variational, state-specific, or multistate approaches, in which ground and excited states are treated on an equal footing by optimizing them explicitly within a common framework.\cite{Ivanov2006, Kossoski2023, Saade2024, Damour2024}
In such approaches, excitation energies are obtained as energy differences between independently or jointly optimized states, and transition properties are evaluated directly from the transition moments between resulting wave functions.
Such formulations may offer a more balanced description of electronic structure and are particularly well-suited for challenging scenarios such as conical intersections and strongly correlated systems.
%
Both of these routes are well-developed within a classical quantum-chemistry context, but there is a growing interest in also adopting quantum computing as a tool for electronic structure theory.
This is motivated by the fact that (ideal) quantum computers naturally and efficiently operate on an exponentially large Hilbert space, and that an exact classical treatment of the many-electron wavefunction similarly requires exponential resources.
This has led to the expectation that quantum algorithms may provide a route to overcoming the exponential bottlenecks inherent in classical approaches to strongly correlated electronic structure problems.
In the long term, these challenges are expected to be addressed by fault-tolerant quantum algorithms, such as quantum phase estimation and related Hamiltonian simulation approaches.
However, current and near-term quantum hardware remains limited by noise and decoherence, which restricts the depth and complexity of circuits that can be reliably executed, thereby precluding the practical realization of such fault-tolerant algorithms.
Thus, in quantum chemistry, this potential has so far been explored primarily through variational hybrid quantum–classical algorithms, or at least using that as a subroutine, most notably the variational quantum eigensolver (VQE)\cite{Peruzzo2014-rx}.
In these approaches, a parameterized quantum circuit is used to prepare a trial wave function, while a classical optimizer iteratively updates the parameters to minimize the measured energy.

LR and equation of motion (EOM) formulations for excited states have been extensively explored in the quantum computing context, including in previous work by some of the present authors\cite{Ollitrault2020-jd,Kumar2023-nf,Asthana2023-eu,Ziems2024-ew,Ziems2024-mt,Jensen2024-hs}.
Although these methods overall appear as a promising direction, LR/EOM methods have been shown to exhibit several limitations in certain regimes, notably from noise sensitivity when the parameterization introduces a metric matrix,\cite{Kjellgren2024-nm,Kwao2026-gd}.
Even for parameterizations that avoid an explicit metric, the resulting excitation energies may depend on parameters that are redundant with respect to the ground-state wave function.\cite{Kjellgren2025-cj}
Furthermore, it is well known that linear-response approaches become inadequate in the vicinity of conical intersections.\cite{Ziems2024-ew,Taylor2024-ji,Taylor2023-tg}
Thus, it is worthwhile to consider the performance of alternative, explicitly state-targeting methods.

Excited states can be targeted through the variance-VQE\cite{Zhang2022-cw} or different formulations of state-averaged VQE, with the two most common approaches being subspace-search VQE (SS-VQE)\cite{Nakanishi2019-ng} and multistate contracted VQE (MC-VQE)\cite{Parrish2019-fi}.
Grimsley and Evangelista found the MC-VQE approach to be a promising alternative to self-consistent EOM.\cite{Grimsley2025-ie}
The state-averaged VQE approaches have also been extended to orbital-optimized versions.\cite{yalouz2021state,Yalouz2022-xz}
In this work, we build upon these developments by extending the orbital-optimized state-averaged VQE framework to a spin-adapted formulation, thereby ensuring the correct spin symmetry of all included states. We focus on the multistate contracted state-resolution,\cite{Parrish2019-fi}, which reduces circuit complexity at the expense of an increased number of measured expectation values during state resolution and subsequent property evaluations.
Importantly, assessing the practical viability of such approaches requires going beyond noiseless simulations and demonstrating their performance under realistic device conditions. To this end, we consider key implementation aspects, including explicit circuit constructions and the incorporation of quantum error mitigation techniques, and perform absorption spectra calculations within the MC-VQE framework on real quantum hardware.

\section{Theory}

\subsection{Multistate Contracted VQE}

In the multistate contracted VQE (MC-VQE)\cite{Parrish2019-fi} model, multiple electronic states are optimized at once using the VQE algorithm.
In the MC-VQE framework, each state is constructed using a two-part circuit: a fixed, parameter-free state-preparation circuit that initializes one of the orthogonal basis states, followed by a parameterized correlator circuit that introduces electron correlation through a set of variational parameters shared across all states.
The optimization procedure targets the state-averaged energy, which is given as
\begin{align}
    E^\text{SA}_0 &= \min_{\boldsymbol{\theta}}\frac{1}{N_\text{states}}\sum_IE_I(\boldsymbol{\theta})\\
    &= \min_{\boldsymbol{\theta}}\frac{1}{N_\text{states}}\sum_I^{N_\text{states}}\left<I\left|\boldsymbol{U}^\dagger(\boldsymbol{\theta})\hat{H}\boldsymbol{U}(\boldsymbol{\theta})\right|I\right>
    \label{eq:ESA}
\end{align}
Here, $\left|I\right>$ are the reference states, which are taken to be orthogonal and to transform according to a definite spin symmetry.
Under a unitary parameterization, the optimization of the state-averaged energy reduces to minimizing the trace of the Hamiltonian within the subspace, and does not by itself provide state-resolved eigenstates, necessitating a subsequent diagonalization within the subspace.
This inter-multistate optimization (also termed \textit{state-resolution)} can be performed using a variety of approaches and differentiates the methods mentioned above. We focus on the approach by Parrish et al.\cite{Parrish2019-fi} using a classical diagonalization of the multistate Hamiltonian,
\begin{equation}
    \boldsymbol{H}^\text{MS}\boldsymbol{V}_{\!\!I} = E_I\boldsymbol{V}_{\!\!I}
\end{equation}
where the multistate Hamiltonian matrix elements are defined as,
\begin{equation}
    H^\text{MS}_{IJ} = \left<I\left|\boldsymbol{U}^\dagger(\boldsymbol{\theta}^\text{opt})\hat{H}\boldsymbol{U}(\boldsymbol{\theta}^\text{opt})\right|J\right>
    \label{eq:HAM_MS}.
\end{equation}
Here, $\boldsymbol{\theta}^\text{opt}$ are the parameters found using Eq. (\ref{eq:ESA}).
It should be noted that the Hamiltonian is hermitian ($H^\text{MS}_{IJ}=\left(H^\text{MS}_{JI}\right)^*$), such that only the upper (or lower) triangular part of the matrix needs to be evaluated.
After performing the classical diagonalization of the multistate Hamiltonian matrix, we obtain the vectors $\boldsymbol{V}_{\!\!I}$, which contain the expansion coefficients of the eigenstates in the reference-state basis. 
These are used to express the final wave functions in the form,
\begin{equation}
    \left|\Psi_I\right> = \boldsymbol{U}(\boldsymbol{\theta}^\text{opt})\sum_JV_{JI}\left|J\right>
\end{equation}
With $V_{JI}$ being the column vectors of $\boldsymbol{V}_I$.

\subsection{Unitary Product State}
In variational quantum algorithms, a variety of ansätze have been proposed for state preparation, including hardware-efficient ansätze (HEA)\cite{Kandala2017-nj}, unitary coupled cluster (UCC)\cite{Taube2006-dv,Peruzzo2014-rx} approaches, or tiled unitary product states ansätze\cite{Burton2024-pn,Anselmetti2021-ra}.

When targeting quantum hardware, a unitary product state can parameterize the MC-VQE.
In this work, we restrict ourselves to UPS-type wave functions, which are written in the following form,
\begin{equation}
    \left|\text{UPS}\right> = \prod_i\boldsymbol{U}_i(\theta_i)\left|I\right>
\end{equation}
with $\boldsymbol{U}_i(\theta_i)$ being a parameterized unitary operator.
The exact eigenstates are eigenfunctions of the total spin operator $S^2$. Maintaining this symmetry can be achieved either by introducing spin penalty terms or by constructing a spin-adapted operator pool. In this work, we take the latter approach and restrict ourselves to operators that preserve this symmetry. 
Recent advances have made spin-adapted double excitation operators possible for both state-vector simulators\cite{Kjellgren2025-pb, Magoulas2025-wb} and circuit representations.\cite{Magoulas2025-nf,Jain2026-tt}
This allows for spin-adapted versions of ans\"atze like factorized UCC, ADAPT\cite{Grimsley2019-yc}, etc.
There also exist spin-adapted ans\"atze that utilize a pool of excitation operators that avoid these more complicated spin-adapted double excitations.
These ans\"atze make use of only singlet spin-adapted single excitation operators and pair-double excitation operators.
These types of excitations are enough to construct the ans\"atze tUPS\cite{Burton2024-pn} and k-UpCCGSD.\cite{Lee2019-ka}
A spin-adapted fermionic ADAPT ansatz can also be constructed using singles and pair-doubles, but this type of ADAPT is difficult to converge\cite{Burton2023-zq} and will not be considered in this work.

Having described the functional form of the quantum circuit, we next turn to the problem of determining the optimal variational parameters.
One possible strategy for optimizing the circuit parameters is to use gradient-based methods, which use gradient information to guide efficient parameter updates during the optimization. The gradient of a UPS wave function can be found using the parameter-shift rule\cite{Wierichs2022-ta}. Since the total energy is the average energy of all of the states, the total gradient is the average gradient over all states.
\begin{align}
    \frac{\partial E^\text{SA}}{\partial \theta_{i}} &= \frac{1}{N_\text{states}}\sum_I^{N_\text{states}}\frac{\partial E_I}{\partial\theta_i}\\
    &= \frac{1}{N_\text{states}}\sum_I^{N_\text{states}}\frac{\partial\left<I\left|\boldsymbol{U}^\dagger(\boldsymbol{\theta})\hat{H}\boldsymbol{U}(\boldsymbol{\theta})\right|I\right>}{\partial\theta_{i}}
\end{align}
Thus, the parameter-shift rule can be used $N_\text{states}$ times to get the gradient of the multistate contracted UPS wave function.
Equivalent considerations can be made to use the gradient-free RotoSolve\cite{Nakanishi2020-mg,Ostaszewski2021-oj,Jager2025-vo,Rossi2026-sd} type optimizers for the multistate contracted UPS wave function.

\subsection{Orbital Optimized MC-VQE}

The MC-VQE can be extended to include orbital optimization, which can be used in combination with the active-space approximation or simply optimize the orbital rotation together with the circuit parameterization.

The orbital rotation parameterization itself can be rewritten as an integral transformation,\cite{Helgaker2013-xk}
\begin{align}
    h_{pq}\left(\boldsymbol{\kappa}\right) &= \sum_{p'q'} \left[\mathrm{e}^{\boldsymbol{\kappa}}\right]_{q'q}h_{p'q'}\left[\mathrm{e}^{-\boldsymbol{\kappa}}\right]_{p'p}\label{eq:int1e_kappa}\\
    g_{pqrs}\left(\boldsymbol{\kappa}\right) &= \sum_{p'q'r's'}\left[\mathrm{e}^{\boldsymbol{\kappa}}\right]_{s's}\left[\mathrm{e}^{\boldsymbol{\kappa}}\right]_{q'q}g_{p'q'r's'}\left[\mathrm{e}^{-\boldsymbol{\kappa}}\right]_{p'p}\left[\mathrm{e}^{-\boldsymbol{\kappa}}\right]_{r'r}\label{eq:int2e_kappa}
\end{align}
making it a purely classic cost.
The minimization of the energy is now over the circuit parameters, $\boldsymbol{\theta}$, and orbital rotation parameters, $\boldsymbol{\kappa}$,
\begin{equation}
    E^\text{SA}_0 = \min_{\boldsymbol{\theta},\boldsymbol{\kappa}}\frac{1}{N_\text{states}}\sum_I^{N_\text{states}}\left<I\left|\boldsymbol{U}^\dagger(\boldsymbol{\theta})\hat{H}(\boldsymbol{\kappa})\boldsymbol{U}(\boldsymbol{\theta})\right|I\right>
    \label{eq:ESA_orb}
\end{equation}
For gradient-based solvers, the gradient of the energy is needed,
\begin{equation}
    \frac{\partial E^\text{SA}}{\partial \kappa_{mn}} = \frac{1}{N_\text{states}}\sum_I^{N_\text{states}}\frac{\partial\left<I\left|\boldsymbol{U}^\dagger(\boldsymbol{\theta})\hat{H}(\boldsymbol{\kappa})\boldsymbol{U}(\boldsymbol{\theta})\right|I\right>}{\partial\kappa_{mn}}
\end{equation}
For the state-specific case, the orbital rotation gradient is given as,
\begin{align}
    \frac{\partial\left<I\left|\boldsymbol{U}^\dagger(\boldsymbol{\theta})\hat{H}(\boldsymbol{\kappa})\boldsymbol{U}(\boldsymbol{\theta})\right|I\right>}{\partial\kappa_{mn}} &= 2\sum_p \left(h_{np}D^{[1]}_{mp} + h_{pm}D^{[1]}_{pn}\right)\\
    &\nonumber\quad+ \sum_{pqr}\left(g_{npqr}D^{[2]}_{mpqr} - g_{pmqr}D^{[2]}_{pnqr} - g_{mpqr}D^{[2]}_{npqr} + g_{pnqr}D^{[2]}_{pmqr}\right)
\end{align}
Hence, the state-averaged orbital gradient takes the form,
\begin{align}
    \frac{\partial E^\text{SA}}{\partial \kappa_{mn}} &= 2\sum_p \left(h_{np}D^{\text{SA},[1]}_{mp} + h_{pm}D^{\text{SA},[1]}_{pn}\right)\\
    &\nonumber\quad+ \sum_{pqr}\left(g_{npqr}D^{\text{SA},[2]}_{mpqr} - g_{pmqr}D^{\text{SA},[2]}_{pnqr} - g_{mpqr}D^{\text{SA},[2]}_{npqr} + g_{pnqr}D^{\text{SA},[2]}_{pmqr}\right)
\end{align}
with $D^{\text{SA},[1]}_{pq}$ and $D^{\text{SA},[2]}_{pqrs}$ being the state-averaged 1-RDM and 2-RDM respectively,
\begin{align}
    D^{\text{SA},[1]}_{pq} &= \frac{1}{N_\text{states}}\sum_I^{N_\text{states}}\left<I\left|\boldsymbol{U}^\dagger(\boldsymbol{\theta})\hat{E}_{pq}\boldsymbol{U}(\boldsymbol{\theta})\right|I\right>\\
    &= \frac{1}{N_\text{states}}\sum_I^{N_\text{states}}D^{I,[1]}_{pq}\\
    D^{\text{SA},[2]}_{pqrs} &= \frac{1}{N_\text{states}}\sum_I^{N_\text{states}}\left<I\left|\boldsymbol{U}^\dagger(\boldsymbol{\theta})\hat{e}_{pqrs}\boldsymbol{U}(\boldsymbol{\theta})\right|I\right>\\
    &= \frac{1}{N_\text{states}}\sum_I^{N_\text{states}}D^{I,[2]}_{pqrs}
\end{align}
For full-space calculations, adding orbital rotations corresponds to applying generalized singles just before the Hamiltonian,
\begin{equation}
    E(\boldsymbol{\theta},\boldsymbol{\kappa}) = \left<I\left|\boldsymbol{U}^\dagger(\boldsymbol{\theta})\boldsymbol{U}^\dagger(\boldsymbol{\kappa})\hat{H}\boldsymbol{U}(\boldsymbol{\kappa})\boldsymbol{U}(\boldsymbol{\theta})\right|I\right>
\end{equation}
with
\begin{equation}
    \boldsymbol{U}(\boldsymbol{\kappa}) = \exp\left(\sum_{p>q}\kappa_{pq}\hat{\sigma}_{pq}^\text{SA}\right)~.
\end{equation}
Multiplying the orbital rotations and the generalised singles together gives,
\begin{align}
    \boldsymbol{U}(\boldsymbol{\kappa})\boldsymbol{U}(\theta_{rs}) &= \exp\left(\sum_{p>q}\kappa_{pq}\hat{\sigma}_{pq}^\text{SA}\right)\exp\left(\theta_{rs}\hat{\sigma}_{rs}^\text{SA}\right)\\
    &= \exp\left(\sum_{p>q}\kappa_{pq}\hat{\sigma}_{pq}^\text{SA} + \theta_{rs}\hat{\sigma}_{rs}^\text{SA} + \left[\sum_{p>q}\kappa_{pq}\hat{\sigma}_{pq}^\text{SA},\theta_{rs}\hat{\sigma}_{rs}^\text{SA}\right] + ...\right)
\end{align}
Since commutators between single excitation operators can only give new single excitation operators, this reduces to
\begin{equation}
    \boldsymbol{U}(\boldsymbol{\kappa})\boldsymbol{U}(\theta_{rs}) = \exp\left(\sum_{p>q}\kappa'_{pq}\hat{\sigma}_{pq}^\text{SA}\right) = \boldsymbol{U}(\boldsymbol{\kappa}')~.
\end{equation}
Thus, any single excitation circuit parameter next to the orbital parameterization is redundant and can be removed from the circuit.

\subsection{Excitation Energies and Oscillator Strengths}

Since in the SA-VQE framework we obtain explicit circuits for each state, any state-specific or inter-state property can be readily obtained. Here, we focus on the excitation energy and oscialltor strength.
The excitation energies can be determined as the energy difference between two states,
\begin{equation}
    \varepsilon_{KL} = \left<\Psi_L\left|\hat{H}\right|\Psi_L\right> - \left<\Psi_K\left|\hat{H}\right|\Psi_K\right>
    \label{eq:exc_energy}
\end{equation}
The oscillator strengths in the length gauge can be determined as,
\begin{equation}
    f_{KL} = \frac{2}{3} \frac{m_e}{\hbar^2} \varepsilon_{LK}\left|\left<\Psi_L\left|\hat{\boldsymbol{\mu}}\right|\Psi_K\right>\right|^2
    \label{eq:osc_str}
\end{equation}
To obtain a simulated absorption spectrum, the discrete excitation lines are broadened using a Gaussian convolution\cite{Barone2011-eo},
\begin{equation}
S_L(\omega) =
\frac{e^2 \pi N_A}{2 \ln(10)\,\epsilon_0 m_e c}
\sum_K f_{KL} \frac{\omega}{\varepsilon_{LK}}
g(\omega - \varepsilon_{LK}; \sigma),
\label{eq:spectrum}
\end{equation}%
where $\sigma$ is a phenomenological broadening parameter and 
\begin{equation}
g(\Delta\omega; \sigma)
=
\frac{1}{\sigma\sqrt{2\pi}}
\exp\!\left(-\frac{\Delta\omega^2}{2\sigma^2}\right).
\end{equation}

\subsection{Matrix Elements}

In the construction of the multistate Hamiltonian, Eq. (\ref{eq:HAM_MS}) and the excitation energies, Eq. (\ref{eq:exc_energy}), matrix elements of the following type need to be evaluated,
\begin{equation}
    M_{IJ}=\left<I\left|\boldsymbol{U}^\dagger\hat{O}_\text{H}\boldsymbol{U}\right|J\right>\label{eq:MIJ},
\end{equation}
where $\hat{O}_\text{H}$ is some Hermitian operator.
To evaluate these matrix elements, we use the technique proposed by Parrish et al\cite{Parrish2019-fi} and Nakanishi et al\cite{Nakanishi2019-ng},
using a superposition decomposition of $\left<I\right|$ and $\left|J\right>$,
\begin{align}
     \Re(M_{IJ}) &= \left<\frac{1}{\sqrt{2}}\left(I+J\right)\left|\boldsymbol{U}^\dagger\hat{O}_\text{H}\boldsymbol{U}\right|\frac{1}{\sqrt{2}}\left(I+J\right)\right>\\
     &\nonumber\quad-\frac{1}{2}\left<I\left|\boldsymbol{U}^\dagger\hat{O}_\text{H}\boldsymbol{U}\right|I\right> - \frac{1}{2}\left<J\left|\boldsymbol{U}^\dagger\hat{O}_\text{H}\boldsymbol{U}\right|J\right>\label{eq:MIJ}
\end{align}
As written, Eq. \eqref{eq:MIJ} may require relatively complicated state-preparation circuits in cases where $\left|I\right>$ and $\left|J\right>$ contain multiple determinants. 
Suppose that the states are expanded as a linear combination of determinants as
\begin{equation}
    \left|I\right> = \sum_i c_i \left|\text{det}_i\right> \label{eq:csf}
\end{equation}
with $c_i$ being the weight of uncontracted  determinants $\left|\text{det}_i\right>$ in the state $\left|I\right>$.
Inserting Eq.~\ref{eq:csf} in Eq.~\ref{eq:MIJ} we find
\begin{equation}
    M_{IJ} = \sum_{ij} c_ic_j\left<\text{det}_i\left|\boldsymbol{U}^\dagger\hat{O}_\text{H}\boldsymbol{U}\right|\text{det}_j\right>\label{eq:exp_val}.
\end{equation}
Thus, we can, as an alternative to creating deep state preparation circuits, consider instead making many, but much simpler, state preparation circuits. In this case, each term is evaluated as
\begin{align}
    \Re\left(\left<\text{det}_i\left|\boldsymbol{U}^\dagger\hat{O}_\text{H}\boldsymbol{U}\right|\text{det}_j\right>\right) &= \left<\frac{1}{\sqrt{2}}\left(\text{det}_i+\text{det}_j\right)\left|\boldsymbol{U}^\dagger\hat{O}_\text{H}\boldsymbol{U}\right|\frac{1}{\sqrt{2}}\left(\text{det}_i+\text{det}_j\right)\right>\\
    &\nonumber\quad - \frac{1}{2}\left<\text{det}_i\left|\boldsymbol{U}^\dagger\hat{O}_\text{H}\boldsymbol{U}\right|\text{det}_i\right> - \frac{1}{2}\left<\text{det}_j\left|\boldsymbol{U}^\dagger\hat{O}_\text{H}\boldsymbol{U}\right|\text{det}_j\right> \label{eq:exp_val_trick}
\end{align}
With this approach, the most complicated state-preparation circuits are a superposition between two determinants, which reduces the number of needed CNOT gates and thus reduces the impact of hardware noise.
We note that this reduction of CNOT gates comes with the downside of more expectation value measurements.
In this work, we prioritize hardware noise reduction over the increase in the number of expectation values, as hardware noise is a limiting factor for current quantum devices.

In the calculation of the oscillator strength, Eq. (\ref{eq:osc_str}), matrix elements calculated over the state-resolved wavefunctions are needed.
These matrix elements take the form,
\begin{align}
    M_{\Psi_I\Psi_J} &= \left<\Psi_I\left|\hat{O}_{\mathrm{H}}\right|\Psi_J\right>\\
    &= \sum_{KL}V_{KI}V_{LJ}\left<K\left|\boldsymbol{U}^\dagger\hat{O}_{\mathrm{H}}\boldsymbol{U} \right|L\right> \\ 
    &= \sum_{KL}\sum_{kl}  V_{KI}V_{LJ} c_k c_l \left<\mathrm{det}_k\left|\boldsymbol{U}^\dagger\hat{O}_{\mathrm{H}}\boldsymbol{U}\right|\mathrm{det}_l\right>
\end{align}
thus requiring the calculation of the full-subspace matrix representation of the operator. 
We note that the evaluation of some matrix elements can be screened out by considering the magnitude of the coefficients $|V_{KI}V_{LJ}c_kc_l|$ -- when these are smaller than a user-defined threshold, the terms can safely be neglected.

\section{Computational Details}

All calculations were run using the \texttt{SlowQuant}\cite{slowquant} package.
Ideal state-vector simulations were carried out with the internal simulator from \texttt{SlowQuant}, whereas calculations including shot noise and simulated quantum device noise were performed using \texttt{Qiskit Aer}, and runs on real quantum hardware used \texttt{Qiskit}\cite{qiskit2024} as a backend. 
The wave function optimizations were performed using BFGS through the \texttt{SciPy}\cite{2020SciPy-NMeth} library.
All states included in the state-averaged wavefunction were picked on the basis of the diagonal elements of the singlet CSF basis CISD Hamiltonian.
The chosen states were those that resulted in the lowest diagonal elements.
Initial Hartree-Fock calculations were performed using \texttt{PySCF}.\cite{Sun2015-pg,Sun2018-ih,Sun2020-cl}

We carried out full-space ADAPT\cite{Grimsley2025-ie,Grimsley2019-yc} and oo-ADAPT\cite{Fitzpatrick2024-ar} calculations on LiH ($R_\text{LiH}=1.671$ Å), H$_2$O ($R_\text{OH}=0.977$ Å, $\theta_\text{HOH}=109.00^\mathrm{o}$) and NH$_3$ ($R_\text{NH}=1.012$ Å, $\theta_\text{HNH}=106.67^\mathrm{o}$) with the STO-3G\cite{Hehre1969-gq,Hehre1970-qw} basis set.
The reference CSFs for the different states can be seen in Appendix Table \ref{tab:lih_ref}-\ref{tab:nh3_ref}.
The operator pool was generalised spin-adapted singles and doubles for the ADAPT calculations and generalised spin-adapted doubles for the oo-ADAPT calculations.
For these calculations, we also obtained reference state-averaged FCI energies using \texttt{SlowQuant}.

Finally, we carried out a series of calculations on formaldehyde/cc-pVDZ\cite{Dunning1989-uj} ($R_\text{CO}=1.202$ Å, $R_\text{CH}=1.298$ Å, $\theta_\text{HCH}=116.44^\mathrm{o}$) and H$_3^+$/aug-cc-pVTZ\cite{Dunning1989-uj,De_Jong2001-bw,Kendall1992-lx} ($R_\text{HH}=0.900$ Å), both targeting real quantum hardware.
In these calculations, we adopted a oo-tUPS ansatz with 1 layer, with the last layer of singles removed, and used a (2e, 3o) active space.
For the calculations on formaldehyde, four states were included in the state-averaging, and the quantum device \texttt{ibm\_marrakesh} was used as the hardware backend.
For the calculations on  H$_3^+$, three states were included in the state-averaging, and the quantum device \texttt{ibm\_aachen} was used as the hardware backend.
For both formaldehyde and H$_3^+$, calculations used 50000 shots per measured Pauli-string, and the number of Pauli-strings was reduced using qubit-wise commutativity and using Pauli-saving. The latter was extended from its previous implementation\cite{Ziems2024-mt} to save Pauli strings per unique reference for single and multi-determinant reference states. 

\subsection{Error Mitigation}
The presence of noise on real quantum devices introduces systematic errors in measured quantities, making error mitigation necessary to obtain useful results.
In this work, we have considered the ansatz-based readout and gate error mitigation, $M_\text{ansatz0}$, as introduced in Ziems et al.\cite{Ziems2024-mt}, where mitigated probability vectors, $\boldsymbol{p}_\text{mitigated}$, are obtained from raw probability vectors, $\boldsymbol{p}_\text{raw}$, as,
\begin{equation}
    \boldsymbol{p}_\text{mitigated}=\boldsymbol{M}_0^{-1}\boldsymbol{p}_\text{raw}
\end{equation}
The $M_\text{ansatz0}$ technique can be seen as an extension of the Readout Error Mitigation (REM) technique that corrects for some amount of circuit errors in addition to read-out errors.
Similar to REM, the methods build a confusion matrix ($\boldsymbol{M}_0$), whose elements $[\boldsymbol{M}_0]_{ij}$ represent the probability of measuring bitstring $i$ given that bitstring $j$ was prepared.
In the $M_\text{ansatz0}$ approach, this construction is extended by prepending the ansatz circuit before the application of the $X$ gates used to prepare the computational basis states, with the ansatz parameters chosen such that it ideally executes an identity operation. For the ansatze we consider, this condition is achieved by setting all $\theta=0$. 
In this way, the resulting confusion matrix captures not only readout errors but also a subset of gate-level errors introduced by the ansatz circuit, allowing these effects to be partially mitigated in the corrected probability distributions.

This error mitigation has so far only been employed for single-reference calculations. However, for oo-MC-VQE, expectation values can have different reference states, Eq. (\ref{eq:exp_val}), and crucially superposition of reference states, Eq. (\ref{eq:exp_val_trick}). The latter needs entangling gates to be constructed and thus incur additional errors.
This presents two options for employing the $M_0$ error-mitigation.
The first option is to make the error-mitigation matrix for only the ansatz circuit, and ignore the error from the state-preparation circuit, which we will refer to as $M_0$ in this work.
The second option is to include the entangler of the multi-determinant reference state in the error characterisation, meaning to calculate an error-mitigation matrix for all the possible unique multi-determinant reference states.
We refer to this extension of the method as $M_{0^+}$.
Although the employed error-mitigation in this work scales exponentially in the number of qubits, we note that a scalable version of the $M_0$ error-mitigation has been developed for tiled ans\"atze.\cite{Rasmussen2025-zu}

All calculations on quantum hardware and simulated quantum hardware also employed post-selection for the correct number of $\alpha$ and $\beta$ electrons for Pauli strings in the computational basis.

\section{Results}

\subsection{Ansatz size\label{sec:ansatz_size}}

When targeting multiple states simultaneously in the VQE optimization, one may expect that the expressability of the ansatz would need to be larger than the expressability of an ansatz targeting just the ground state.
To investigate how the ansatz size increases with the number of states, we consider three model systems: LiH (4e, 6o), H$_2$O (10e, 7o), and NH$_3$ (10e, 8o).
\begin{figure}[H]
    \centering
    \includegraphics[width=1.0\linewidth]{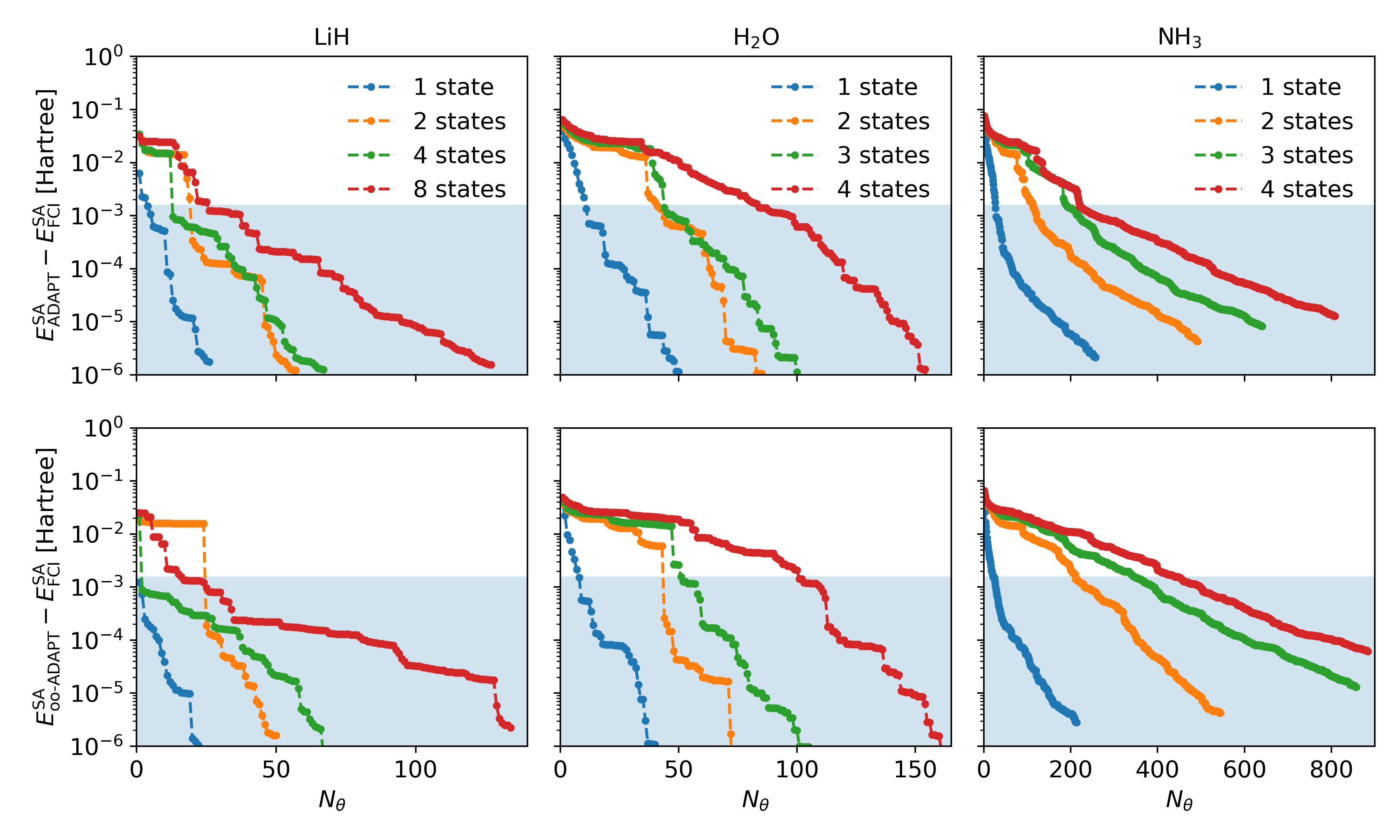}
    \caption{State-averaged ADAPT (top row) and state-averaged oo-ADAPT (bottom row) calculations for LiH H$_2$O and NH$_3$ with different number of states.
    The blue area signifies error below $1.6$ mH.}
    \label{fig:adapt}
\end{figure}
Fig. \ref{fig:adapt} shows the error in the energy with respect to the state-averaged FCI energy for state-averaged ADAPT (top row) and oo-ADAPT (bottom row) for selected systems.
It is evident that the number of required parameters in the circuit increases with the number of states being included in the state-averaged wavefunction.
Specifically, we find that the number of required parameters is roughly linear in the number of states.
For example, requiring an error of at most $10^{-4}$ Hatree with ADAPT, LiH requires 11, 18, 9, and 8 parameters per state for 1, 2, 4, and 8 states, respectively.
H$_2$O requires 27, 32, 24, 30 parameters per state for 1, 2, 3, and 4 states, respectively.
NH$_3$ requires 65, 120, 122, and 132 parameters per state for 1, 2, 3, and 4 states, respectively.

Comparing the number of parameters required between ADAPT and oo-ADAPT, it is apparent that folding the single excitations into the classic orbital optimisation has only a minor effect, or even a detrimental effect, as can be seen for NH$_3$ with 3 (green lines) or 4 (red lines) states.
\begin{figure}[H]
    \centering
    \includegraphics[width=0.5\linewidth]{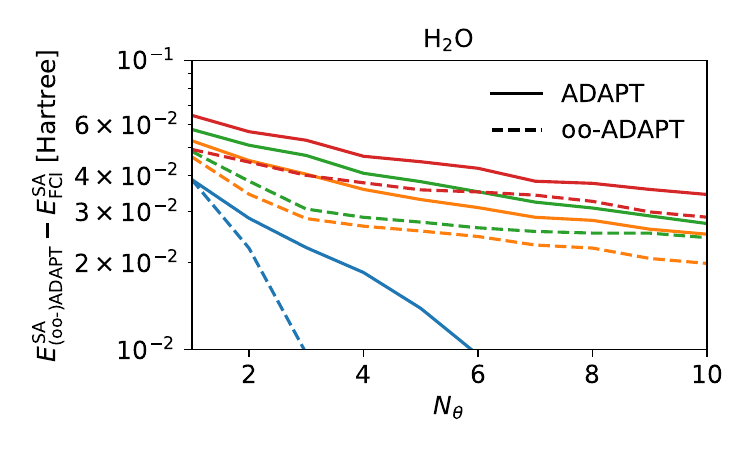}
    \caption{Convergence of state-averaged ADAPT for the first 10 iterations for H$_2$O.
    The colors match that of Fig. \ref{fig:adapt}, blue being 1 state, orange being 2 states, green being 3 states, and, red being 4 states.}
    \label{fig:h2o_zoom}
\end{figure}
The effect of performing orbital-optimization seems to be largest for small expansions, as exemplified in Fig. \ref{fig:h2o_zoom}, which shows the first few iterations of ADAPT and oo-ADAPT for H$_2$O.
Early in the ADAPT procedure, including orbital optimization always leads to smaller energy errors, but as tighter convergence is approached, performing orbital optimization has little effect on the number of parameters in the ADAPT ansatz.

\subsection{Quantum experiments}
To assess the performance of the proposed method under realistic conditions, we now consider its execution on real quantum hardware.
We perform quantum experiments using pre-optimized circuits, meaning the spin-adapted oo-SA-VQE is performed on a noiseless state-vector simulator, and the state-resolution Eq. (\ref{eq:HAM_MS}), the excitation energies Eq. (\ref{eq:exc_energy}), and oscillator strengths Eq. (\ref{eq:osc_str}) are calculated on (emulated) hardware. This corresponds to the standard proof-of-concept VQE approach, where the circuits are optimized classically, and the energy is simulated on the quantum device. 

As examples, we consider two systems. The first system being formaldehyde with 2 electrons in 3 spatial orbitals, using four references states, $\left|110000\right>$, $\frac{1}{\sqrt{2}}\left(\left|100100\right> - \left|011000\right>\right)$, $\frac{1}{\sqrt{2}}\left(\left|100001\right> - \left|010010\right>\right)$, and, $\left|001100\right>$.
And, the second system being H$_3^+$, with 2 electrons in 3 spatial orbitals, using three references states, $\left|110000\right>$, $\frac{1}{\sqrt{2}}\left(\left|100100\right> - \left|011000\right>\right)$, $\frac{1}{\sqrt{2}}\left(\left|100001\right> - \left|010010\right>\right)$.
\begin{figure}[H]
    \centering
    \includegraphics[width=0.5\linewidth]{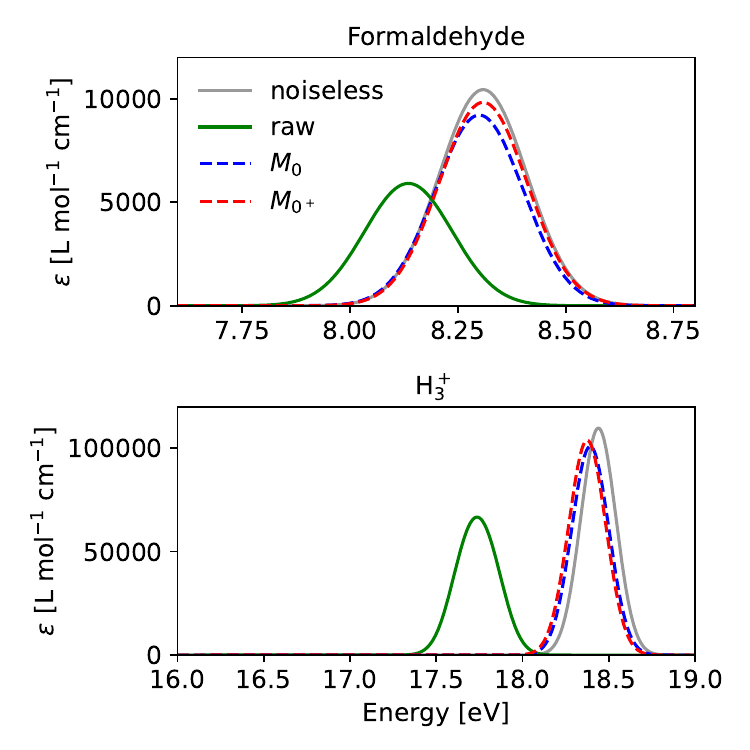}
    \caption{Electronic spectrum of formaldehyde (top row) using simulated noise based on \texttt{ibm\_marrakesh}, and electronic spectrum of H$_3^+$ (bottom row) using simulated noise based on \texttt{ibm\_aachen}.}
    \label{fig:sim_noise}
\end{figure}
First, we perform the simulation on emulated hardware using IBM's noise models. In Fig. \ref{fig:sim_noise}, the calculated spectra for formaldehyde (top plot) and H$_3^+$ (bottom plot) are shown.
The spectra are made by performing a Gaussian convolution with a width of 0.12 eV.
For both of the error mitigation strategies $M_0$ (blue dashed-line) and $M_{0^+}$ (red dashed-line), produce results that are almost identical to the noiseless results (grey line), with $M_{0^+}$ being slightly more accurate, as expected when characterizing both state preparation and ansatz.
When comparing the raw simulated noise results (green lines) between formaldehyde (top plot) and H$_3^+$ (bottom plot), it can be seen that the H$_3^+$ spectrum is much more adversely affected.
For formaldehyde, the peak is shifted $-0.17$ eV ($-6.4$ mHa), whereas for H$_3^+$, it is shifted $-0.70$ eV ($-26$ mHa).
This difference can be explained by considering the condition number of $\boldsymbol{H}^\text{MS}$, as this metric indicates how sensitive the diagonalisation is to noise. 
The condition numbers are found to be $\mathrm{cond}\left(\boldsymbol{H}^\text{MS}_\text{formaldehyde}\right)=1.003$ and $\mathrm{cond}\left(\boldsymbol{H}^\text{MS}_{\text{H}_3^+}\right)=1.284$, which aligns with the observation that the H$_3^+$ spectrum is more adversely affected by noise.
\begin{figure}[H]
    \centering
    \includegraphics[width=0.5\linewidth]{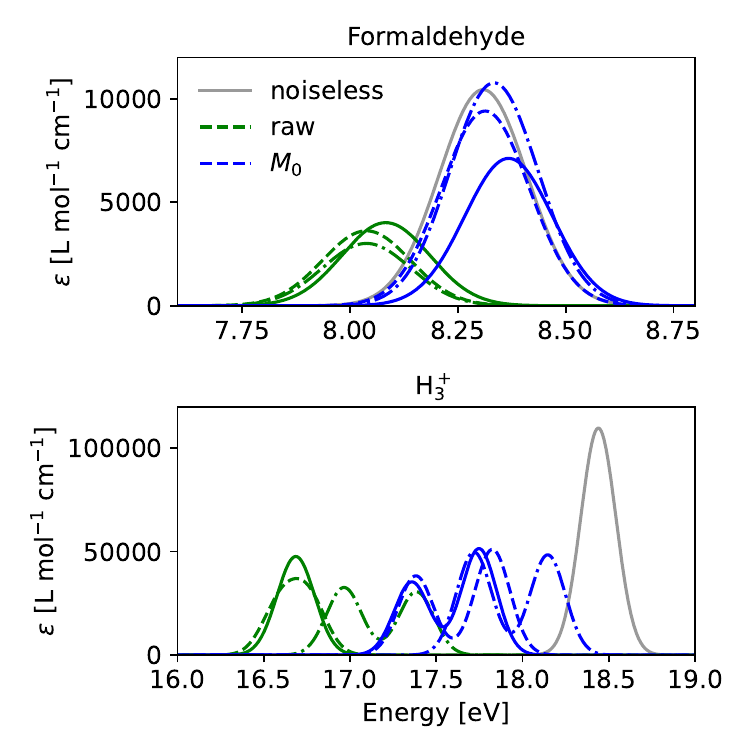}
    \caption{Electronic spectrum of formaldehyde (top row) calculated using the quantum device \texttt{ibm\_marrakesh}, and electronic spectrum of H$_3^+$ (bottom row) using the quantum device \texttt{ibm\_aachen}.
    Both plots include three independent runs, with the different line styles indicating each run.}
    \label{fig:device_spec}
\end{figure}
Finally, we consider the actual quantum hardware experiments. 
In these calculations, we resorted to using the regular $M_0$ error-mitigation instead of the $M_{0^+}$ error-mitigation, since the latter would be prohibitively expensive, and the results on simulated device-noise showed only a minor difference. 

In Fig. \ref{fig:device_spec}, the calculated adsorption spectrum of formaldehyde (top plot) and H$_3^+$ (bottom plot) using quantum devices is shown.
It is clear that the error mitigated results from the real devices (blue dashed lines) are significantly worse than the error mitigated results using simulated noise (Fig. \ref{fig:sim_noise}, blue dashed lines).
This has also been seen in a previous work.\cite{Rasmussen2025-zu}
The calculations on quantum hardware show the same trends as on the simulated devices; the calculated spectrum for H$_3^+$ is more affected by the noise than the calculated spectrum for formaldehyde.
The discrepancy between simulated and real device performance highlights the limitations of current noise models in fully capturing hardware-specific error sources, such as time-dependent noise, crosstalk, and calibration drift. Despite this, the qualitative agreement in trends between simulated and real devices suggests that the proposed methodology remains robust, with the primary features of the spectra preserved under realistic noise conditions. These results indicate that while quantitative accuracy remains challenging on present-day hardware, meaningful chemical insight may still be extracted when combined with appropriate error mitigation techniques.

\section{Conclusion}

In this work, we extend the oo-SA-VQE formalism to a spin-adapted framework, which guarantees the spin state of the reference wave function to be conserved.
It is investigated through ADAPT state-vector simulations how the circuit expressibility requirements increase with the number of states included in the state-averaging.
It is found that doubling the number of states roughly doubles the length of the required circuit.
Further, it is shown that performing orbital optimization together with the ADAPT has a negligible effect on the number of required circuit parameters.

Molecular spectra are calculated with MC-VQE using real quantum hardware.
It is found that molecular spectra can be calculated for some systems, but error-mitigation techniques are necessary to reduce the noise.

\acknowledgments
We acknowledge the support of the Novo Nordisk Foundation (NNF) for the focused research project ``Hybrid Quantum Chemistry on Hybrid Quantum Computers'' (grant number  NNFSA220080996).
K.M.Z. acknowledges financial support from the Royal Society of Chemistry Collaboration grant, C25-1492721325.

\section*{DATA AVAILABILITY}
The data that support the findings of this study are available from the corresponding author upon reasonable request.

\section*{Appendix}

Reference CSFs for LiH, H$_2$O, and NH$_3$.
The energies in the tables are calculated as $\left<\text{CSF}\left|\hat{H}\left(\kappa_\text{HF}\right)\right|\text{CSF}\right>$.

\begin{table}[H]
    \centering
    \begin{tabular}{c|c}
       Energy [Hartree]  &  CSF\\\hline
        -8.80884 & $\left|111100000000\right>$\\
        -8.64620 & $\frac{1}{\sqrt{2}}\left(\left|110110000000\right> - \left|111001000000\right>\right)$ \\
        -8.58640 & $\frac{1}{\sqrt{2}}\left(\left|110100100000\right> - \left|111000010000\right>\right)$ \\
        -8.58640 & $\frac{1}{\sqrt{2}}\left(\left|111000000100\right> - \left|110100001000\right>\right)$ \\
        -8.20356 & $\frac{1}{\sqrt{2}}\left(\left|111000000001\right> - \left|110100000010\right>\right)$\\
        -8.16494 & $\frac{1}{\sqrt{2}}\left(\left|110000100100\right> - \left|110000011000\right>\right)$ \\
        -8.14807 & $\left|110000110000\right>$\\ 
        -8.14807 & $\left|110000001100\right>$
    \end{tabular}
    \caption{Used reference CSFs for LiH calculations.}
    \label{tab:lih_ref}
\end{table}

\begin{table}[H]
    \centering
    \begin{tabular}{c|c}
       Energy [Hartree]  &  CSF\\\hline
        -83.95887 & $\left|111100000000\right>$\\
        -83.50262 & $\frac{1}{\sqrt{2}}\left(\left|11111111100100\right> - \left|11111111011000\right>\right)$ \\
        -83.41635 & $\frac{1}{\sqrt{2}}\left(\left|11111111100001\right> - \left|11111111010010\right>\right)$ \\
        -83.34914 & $\frac{1}{\sqrt{2}}\left(\left|11111110110100\right> - \left|11111101111000\right>\right)$
    \end{tabular}
    \caption{Used reference CSFs for H$_2$O calculations.}
    \label{tab:h2o_ref}
\end{table}

\begin{table}[H]
    \centering
    \begin{tabular}{c|c}
       Energy [Hartree]  &  CSF\\\hline
        -67.40808 & $\left|1111111111000000\right>$\\
        -66.83776  & $\frac{1}{\sqrt{2}}\left(\left|1111111110010000\right> - \left|1111111101100000\right>\right)$\\
        -66.80493  & $\frac{1}{\sqrt{2}}\left(\left|1111111110000100\right> - \left|1111111101001000\right>\right)$\\
        -66.80488  & $\frac{1}{\sqrt{2}}\left(\left|1111111110000001\right> - \left|1111111101000010\right>\right)$
    \end{tabular}
    \caption{Used reference CSFs for NH$_3$ calculations.}
    \label{tab:nh3_ref}
\end{table}

\newpage
\bibliographystyle{unsrt}
\bibliography{literature}

@software{slowquant,
  author = {Erik Kjellgren and Karl Michael Ziems},
  title = {Slow{Q}uant},
  url = {https://github.com/erikkjellgren/SlowQuant/tree/master},
note = {https://github.com/erikkjellgren/SlowQuant/tree/master},
  version = {0.0.0},
  year = {2025},
}

@ARTICLE{Parrish2019-fi,
  title     = "Quantum computation of electronic transitions using a
               variational quantum eigensolver",
  author    = "Parrish, Robert M and Hohenstein, Edward G and McMahon, Peter L
               and Mart{\'\i}nez, Todd J",
  journal   = "Phys. Rev. Lett.",
  publisher = "American Physical Society (APS)",
  volume    =  122,
  number    =  23,
  pages     = "230401",
  month     =  jun,
  year      =  2019,
  copyright = "https://link.aps.org/licenses/aps-default-license",
  language  = "en"
}

@ARTICLE{Burton2024-pn,
  title     = "Accurate and gate-efficient quantum \textit{Ans{\"a}tze} for
               electronic states without adaptive optimization",
  author    = "Burton, Hugh G A",
  journal   = "Phys. Rev. Res.",
  publisher = "American Physical Society (APS)",
  volume    =  6,
  number    =  2,
  month     =  jun,
  year      =  2024,
  copyright = "https://creativecommons.org/licenses/by/4.0/",
  language  = "en"
}

@ARTICLE{Lee2019-ka,
  title     = "Generalized unitary coupled cluster wave functions for quantum
               computation",
  author    = "Lee, Joonho and Huggins, William J and Head-Gordon, Martin and
               Whaley, K Birgitta",
  journal   = "J. Chem. Theory Comput.",
  publisher = "American Chemical Society (ACS)",
  volume    =  15,
  number    =  1,
  pages     = "311--324",
  month     =  jan,
  year      =  2019,
  language  = "en"
}

@misc{qiskit2024,
      title={Quantum computing with {Q}iskit},
      author={Javadi-Abhari, Ali and Treinish, Matthew and Krsulich, Kevin and Wood, Christopher J. and Lishman, Jake and Gacon, Julien and Martiel, Simon and Nation, Paul D. and Bishop, Lev S. and Cross, Andrew W. and Johnson, Blake R. and Gambetta, Jay M.},
      year={2024},
      doi={10.48550/arXiv.2405.08810},
      eprint={2405.08810},
      archivePrefix={arXiv},
      primaryClass={quant-ph},
      journal  = "arXiv",
}

@ARTICLE{Nakanishi2019-ng,
  title     = "Subspace-search variational quantum eigensolver for excited
               states",
  author    = "Nakanishi, Ken M and Mitarai, Kosuke and Fujii, Keisuke",
  journal   = "Phys. Rev. Res.",
  publisher = "American Physical Society (APS)",
  volume    =  1,
  number    =  3,
  month     =  oct,
  year      =  2019,
  copyright = "https://creativecommons.org/licenses/by/4.0/",
  language  = "en"
}

@ARTICLE{Ziems2024-ew,
  title     = "Which options exist for {NISQ-friendly} linear response
               formulations?",
  author    = "Ziems, Karl Michael and Kjellgren, Erik Rosendahl and Reinholdt,
               Peter and Jensen, Phillip W K and Sauer, Stephan P A and
               Kongsted, Jacob and Coriani, Sonia",
  journal   = "J. Chem. Theory Comput.",
  publisher = "American Chemical Society (ACS)",
  volume    =  20,
  number    =  9,
  pages     = "3551--3565",
  month     =  may,
  year      =  2024,
  language  = "en"
}

@ARTICLE{Ziems2024-mt,
  title     = "Understanding and mitigating noise in molecular quantum linear
               response for spectroscopic properties on quantum computers",
  author    = "Ziems, Karl Michael and Kjellgren, Erik Rosendahl and Sauer,
               Stephan P A and Kongsted, Jacob and Coriani, Sonia",
  journal   = "Chem. Sci.",
  publisher = "Royal Society of Chemistry (RSC)",
  volume    =  16,
  number    =  10,
  pages     = "4456--4468",
  month     =  mar,
  year      =  2025,
  copyright = "http://creativecommons.org/licenses/by-nc/3.0/",
  language  = "en"
}

@ARTICLE{Hehre1969-gq,
  title     = "Self-consistent molecular-orbital methods. I. use of Gaussian
               expansions of Slater-type atomic orbitals",
  author    = "Hehre, W J and Stewart, R F and Pople, J A",
  journal   = "J. Chem. Phys.",
  publisher = "AIP Publishing",
  volume    =  51,
  number    =  6,
  pages     = "2657--2664",
  month     =  sep,
  year      =  1969,
  language  = "en"
}

@ARTICLE{Hehre1970-qw,
  title     = "Self-consistent molecular orbital methods. {IV}. Use of Gaussian
               expansions of Slater-type orbitals. Extension to second-row
               molecules",
  author    = "Hehre, W J and Ditchfield, R and Stewart, R F and Pople, J A",
  journal   = "J. Chem. Phys.",
  publisher = "AIP Publishing",
  volume    =  52,
  number    =  5,
  pages     = "2769--2773",
  month     =  mar,
  year      =  1970,
  language  = "en"
}

@ARTICLE{Kjellgren2024-nm,
  title    = "Divergences in classical and quantum linear response and equation
              of motion formulations",
  author   = "Kjellgren, Erik Rosendahl and Reinholdt, Peter and Ziems, Karl
              Michael and Sauer, Stephan P A and Coriani, Sonia and Kongsted,
              Jacob",
  journal  = "J. Chem. Phys.",
  pages ={124112},
  volume   =  161,
  number   =  12,
  month    =  sep,
  year     =  2024,
  language = "en"
}

@ARTICLE{Jensen2024-hs,
  title     = "Quantum equation of motion with orbital optimization for
               computing molecular properties in near-term quantum computing",
  author    = "Jensen, Phillip W K and Kjellgren, Erik Rosendahl and Reinholdt,
               Peter and Ziems, Karl Michael and Coriani, Sonia and Kongsted,
               Jacob and Sauer, Stephan P A",
  journal   = "J. Chem. Theory Comput.",
  publisher = "American Chemical Society (ACS)",
  volume    =  20,
  number    =  9,
  pages     = "3613--3625",
  month     =  may,
  year      =  2024,
  language  = "en"
}

@ARTICLE{2020SciPy-NMeth,
  author  = {Virtanen, Pauli and Gommers, Ralf and Oliphant, Travis E. and
            Haberland, Matt and Reddy, Tyler and Cournapeau, David and
            Burovski, Evgeni and Peterson, Pearu and Weckesser, Warren and
            Bright, Jonathan and {van der Walt}, St{\'e}fan J. and
            Brett, Matthew and Wilson, Joshua and Millman, K. Jarrod and
            Mayorov, Nikolay and Nelson, Andrew R. J. and Jones, Eric and
            Kern, Robert and Larson, Eric and Carey, C J and
            Polat, {\.I}lhan and Feng, Yu and Moore, Eric W. and
            {VanderPlas}, Jake and Laxalde, Denis and Perktold, Josef and
            Cimrman, Robert and Henriksen, Ian and Quintero, E. A. and
            Harris, Charles R. and Archibald, Anne M. and
            Ribeiro, Ant{\^o}nio H. and Pedregosa, Fabian and
            {van Mulbregt}, Paul and {SciPy 1.0 Contributors}},
  title   = {{{SciPy} 1.0: Fundamental Algorithms for Scientific
            Computing in Python}},
  journal = {Nature Methods},
  year    = {2020},
  volume  = {17},
  pages   = {261--272},
  adsurl  = {https://rdcu.be/b08Wh},
  doi     = {10.1038/s41592-019-0686-2},
}

@Book{Helgaker2013-xk,
  author    = {Helgaker, Trygve and J{\o}rgensen, Poul and Olsen, Jeppe},
  title     = {Molecular electronic-structure theory},
  publisher = {John Wiley \& Sons},
  address   = {Nashville, TN},
  month     = feb,
  year      = {2013},
}

@ARTICLE{Grimsley2025-ie,
  title     = "Challenging excited states from adaptive quantum eigensolvers:
               subspace expansions vs. state-averaged strategies",
  author    = "Grimsley, Harper R and Evangelista, Francesco A",
  journal   = "Quantum Sci. Technol.",
  publisher = "IOP Publishing",
  volume    =  10,
  number    =  2,
  pages     = "025003",
  month     =  apr,
  year      =  2025,
  copyright = "https://creativecommons.org/licenses/by/4.0/"
}

@ARTICLE{Peruzzo2014-rx,
  title     = "A variational eigenvalue solver on a photonic quantum processor",
  author    = "Peruzzo, Alberto and McClean, Jarrod and Shadbolt, Peter and
               Yung, Man-Hong and Zhou, Xiao-Qi and Love, Peter J and
               Aspuru-Guzik, Al{\'a}n and O'Brien, Jeremy L",
  journal   = "Nat. Commun.",
  publisher = "Springer Science and Business Media LLC",
  volume    =  5,
  number    =  1,
  pages     = "4213",
  month     =  jul,
  year      =  2014,
  copyright = "https://creativecommons.org/licenses/by-nc-nd/4.0",
  language  = "en"
}

@article{yalouz2021state,
  title={A state-averaged orbital-optimized hybrid quantum--classical algorithm for a democratic description of ground and excited states},
  author={Yalouz, Saad and Senjean, Bruno and G{\"u}nther, Jakob and Buda, Francesco and O’Brien, Thomas E and Visscher, Lucas},
  journal={Quantum Science and Technology},
  volume={6},
  number={2},
  pages={024004},
  year={2021},
  publisher={IOP Publishing}
}

@ARTICLE{Zhang2022-cw,
  title     = "Variational quantum eigensolvers by variance minimization",
  author    = "Zhang, Dan-Bo and Chen, Bin-Lin and Yuan, Zhan-Hao and Yin, Tao",
  journal   = "Chin. Physics B",
  publisher = "IOP Publishing",
  volume    =  31,
  number    =  12,
  pages     = "120301",
  month     =  nov,
  year      =  2022,
  copyright = "https://iopscience.iop.org/page/copyright"
}

@ARTICLE{Kjellgren2025-pb,
  title     = "Exact closed-form expressions for unitary spin-adapted fermionic
               singlet double excitation operators",
  author    = "Kjellgren, Erik Rosendahl and Ziems, Karl Michael and Reinholdt,
               Peter and Sauer, Stephan P A and Coriani, Sonia and Kongsted,
               Jacob",
  journal   = "J. Chem. Phys.",
  publisher = "AIP Publishing",
  volume    =  163,
  number    =  13,
  month     =  oct,
  year      =  2025,
  pages     =  {134115}, 
  language  = "en"
}

@ARTICLE{Magoulas2025-wb,
  title     = "Closed-form expressions for unitaries of spin-adapted fermionic
               operators",
  author    = "Magoulas, Ilias and Evangelista, Francesco A",
  journal   = "Mol. Phys.",
  publisher = "Informa UK Limited",
  number    = "e2534672",
  month     =  aug,
  year      =  2025,
  language  = "en"
}

@ARTICLE{Burton2023-zq,
  title     = "Exact electronic states with shallow quantum circuits from
               global optimisation",
  author    = "Burton, Hugh G A and Marti-Dafcik, Daniel and Tew, David P and
               Wales, David J",
  journal   = "Npj Quantum Inf.",
  publisher = "Springer Science and Business Media LLC",
  volume    =  9,
  number    =  1,
  month     =  jul,
  year      =  2023,
  copyright = "https://creativecommons.org/licenses/by/4.0",
  language  = "en"
}

@ARTICLE{Kjellgren2025-cj,
  title        = "Redundant parameter dependencies in truncated classic and
                  quantum Linear Response and Equation of Motion theory",
  author       = "Kjellgren, Erik Rosendahl and Reinholdt, Peter and Ziems,
                  Karl Michael and Sauer, Stephan P A and Coriani, Sonia and
                  Kongsted, Jacob",
 journal  = {J. Chem. Phys.},
 volume = {163},
 pages  = {134111},
  doi={10.1063/5.0284287},
  year         =  2025,
  primaryClass = "physics.chem-ph",
  eprint       = "2506.06063"
}

@ARTICLE{Dunning1989-uj,
  title     = "Gaussian basis sets for use in correlated molecular
               calculations. I. The atoms boron through neon and hydrogen",
  author    = "Dunning, Jr, Thom H",
  journal   = "J. Chem. Phys.",
  publisher = "AIP Publishing",
  volume    =  90,
  number    =  2,
  pages     = "1007--1023",
  month     =  jan,
  year      =  1989,
  language  = "en"
}

@ARTICLE{Grimsley2019-yc,
  title     = "An adaptive variational algorithm for exact molecular
               simulations on a quantum computer",
  author    = "Grimsley, Harper R and Economou, Sophia E and Barnes, Edwin and
               Mayhall, Nicholas J",
  journal   = "Nat. Commun.",
  publisher = "Springer Science and Business Media LLC",
  volume    =  10,
  number    =  1,
  pages     = "3007",
  month     =  jul,
  year      =  2019,
  copyright = "https://creativecommons.org/licenses/by/4.0",
  language  = "en"
}

@ARTICLE{Fitzpatrick2024-ar,
  title    = "Self-consistent field approach for the variational quantum
              eigensolver: Orbital optimization goes adaptive",
  author   = "Fitzpatrick, Aaron and Nyk{\"a}nen, Anton and Talarico, N Walter
              and Lunghi, Alessandro and Maniscalco, Sabrina and
              Garc{\'\i}a-P{\'e}rez, Guillermo and Knecht, Stefan",
  journal  = "J. Phys. Chem. A",
  volume   =  128,
  number   =  14,
  pages    = "2843--2856",
  month    =  apr,
  year     =  2024,
  language = "en"
}

@ARTICLE{Sun2020-cl,
  title     = "Recent developments in the {PySCF} program package",
  author    = "Sun, Qiming and Zhang, Xing and Banerjee, Samragni and Bao, Peng
               and Barbry, Marc and Blunt, Nick S and Bogdanov, Nikolay A and
               Booth, George H and Chen, Jia and Cui, Zhi-Hao and Eriksen,
               Janus J and Gao, Yang and Guo, Sheng and Hermann, Jan and
               Hermes, Matthew R and Koh, Kevin and Koval, Peter and Lehtola,
               Susi and Li, Zhendong and Liu, Junzi and Mardirossian, Narbe and
               McClain, James D and Motta, Mario and Mussard, Bastien and Pham,
               Hung Q and Pulkin, Artem and Purwanto, Wirawan and Robinson,
               Paul J and Ronca, Enrico and Sayfutyarova, Elvira R and
               Scheurer, Maximilian and Schurkus, Henry F and Smith, James E T
               and Sun, Chong and Sun, Shi-Ning and Upadhyay, Shiv and Wagner,
               Lucas K and Wang, Xiao and White, Alec and Whitfield, James
               Daniel and Williamson, Mark J and Wouters, Sebastian and Yang,
               Jun and Yu, Jason M and Zhu, Tianyu and Berkelbach, Timothy C
               and Sharma, Sandeep and Sokolov, Alexander Yu and Chan, Garnet
               Kin-Lic",
  journal   = "J. Chem. Phys.",
  publisher = "AIP Publishing",
  volume    =  153,
  number    =  2,
  pages     = "024109",
  month     =  jul,
  year      =  2020,
  language  = "en"
}

@ARTICLE{Sun2018-ih,
  title     = "{P}y{SCF}: the Python‐based simulations of chemistry framework",
  author    = "Sun, Qiming and Berkelbach, Timothy C and Blunt, Nick S and
               Booth, George H and Guo, Sheng and Li, Zhendong and Liu, Junzi
               and McClain, James D and Sayfutyarova, Elvira R and Sharma,
               Sandeep and Wouters, Sebastian and Chan, Garnet Kin-Lic",
  journal   = "Wiley Interdiscip. Rev. Comput. Mol. Sci.",
  publisher = "Wiley",
  volume    =  8,
  number    =  1,
  pages     = "e1340",
  month     =  jan,
  year      =  2018,
  copyright = "http://onlinelibrary.wiley.com/termsAndConditions\#vor",
  language  = "en"
}

@ARTICLE{Sun2015-pg,
  title     = "Libcint: An efficient general integral library for Gaussian
               basis functions",
  author    = "Sun, Qiming",
  journal   = "J. Comput. Chem.",
  publisher = "Wiley",
  volume    =  36,
  number    =  22,
  pages     = "1664--1671",
  month     =  aug,
  year      =  2015,
  language  = "en"
}

@ARTICLE{Kwao2026-gd,
  title    = "Generalized eigenvalue problem in subspace-based excited-state
              methods for quantum computers",
  author   = "Kwao, Prince Frederick and Poyyapakkam Sundar, Srivathsan and
              Gupt, Brajesh and Asthana, Ayush",
  journal  = "J. Chem. Theory Comput.",
  month    =  mar,
  year     =  2026,
  language = "en"
}

@ARTICLE{Rasmussen2025-zu,
  title         = "Cost-effective scalable quantum error mitigation for tiled
                   Ans{\"a}tze",
  author        = "Rasmussen, Oskar Graulund Lentz and Kjellgren, Erik and
                   Reinholdt, Peter and Sauer, Stephan P A and Coriani, Sonia
                   and Ziems, Karl Michael and Kongsted, Jacob",
  month         =  nov,
  year          =  2025,
  copyright     = "http://creativecommons.org/licenses/by/4.0/",
  archivePrefix = "arXiv",
  primaryClass  = "quant-ph",
  eprint        = "2511.21236",
  journal  = "arXiv",
}

@ARTICLE{Nakanishi2020-mg,
  title     = "Sequential minimal optimization for quantum-classical hybrid
               algorithms",
  author    = "Nakanishi, Ken M and Fujii, Keisuke and Todo, Synge",
  journal   = "Phys. Rev. Res.",
  publisher = "American Physical Society (APS)",
  volume    =  2,
  number    =  4,
  month     =  oct,
  year      =  2020,
  copyright = "https://creativecommons.org/licenses/by/4.0/",
  language  = "en"
}

@ARTICLE{Ostaszewski2021-oj,
  title     = "Structure optimization for parameterized quantum circuits",
  author    = "Ostaszewski, Mateusz and Grant, Edward and Benedetti, Marcello",
  journal   = "Quantum",
  publisher = "Verein zur Forderung des Open Access Publizierens in den
               Quantenwissenschaften",
  volume    =  5,
  number    =  391,
  pages     = "391",
  month     =  jan,
  year      =  2021,
  language  = "en"
}

@ARTICLE{Jager2025-vo,
  title     = "Fast gradient-free optimization of excitations in variational
               quantum eigensolvers",
  author    = "J{\"a}ger, Jonas and Kaldenbach, Thierry N and Haas, Max and
               Schultheis, Erik",
  journal   = "Commun. Phys.",
  publisher = "Springer Science and Business Media LLC",
  volume    =  8,
  number    =  1,
  pages     = "418",
  month     =  oct,
  year      =  2025,
  copyright = "https://creativecommons.org/licenses/by/4.0",
  language  = "en"
}

@ARTICLE{Yalouz2022-xz,
  title     = "Analytical nonadiabatic couplings and gradients within the
               state-averaged orbital-optimized variational quantum eigensolver",
  author    = "Yalouz, Saad and Koridon, Emiel and Senjean, Bruno and Lasorne,
               Benjamin and Buda, Francesco and Visscher, Lucas",
  journal   = "J. Chem. Theory Comput.",
  publisher = "American Chemical Society (ACS)",
  volume    =  18,
  number    =  2,
  pages     = "776--794",
  month     =  feb,
  year      =  2022,
  copyright = "https://creativecommons.org/licenses/by-nc-nd/4.0/",
  language  = "en"
}

@ARTICLE{De_Jong2001-bw,
  title     = "Parallel {Douglas--Kroll} energy and gradients in {NWChem}:
               Estimating scalar relativistic effects using {Douglas--Kroll}
               contracted basis sets",
  author    = "de Jong, W A and Harrison, R J and Dixon, D A",
  journal   = "J. Chem. Phys.",
  publisher = "AIP Publishing",
  volume    =  114,
  number    =  1,
  pages     = "48--53",
  month     =  jan,
  year      =  2001,
  language  = "en"
}

@ARTICLE{Kendall1992-lx,
  title     = "Electron affinities of the first-row atoms revisited. Systematic
               basis sets and wave functions",
  author    = "Kendall, Rick A and Dunning, Jr, Thom H and Harrison, Robert J",
  journal   = "J. Chem. Phys.",
  publisher = "AIP Publishing",
  volume    =  96,
  number    =  9,
  pages     = "6796--6806",
  month     =  may,
  year      =  1992,
  language  = "en"
}

@ARTICLE{Magoulas2025-nf,
  title         = "Spin-adapted fermionic unitaries: From Lie algebras to
                   compact quantum circuits",
  author        = "Magoulas, Ilias and Evangelista, Francesco A",
  month         =  nov,
  year          =  2025,
  copyright     = "http://arxiv.org/licenses/nonexclusive-distrib/1.0/",
  archivePrefix = "arXiv",
  primaryClass  = "quant-ph",
  eprint        = "2511.13485",
  journal  = "arXiv",
}

@ARTICLE{Jain2026-tt,
  title     = "Exact factorization of unitary transformations with spin-adapted
               generators",
  author    = "Jain, Paarth and Izmaylov, Artur F and Kjellgren, Erik R",
  journal   = "J. Chem. Phys.",
  publisher = "AIP Publishing",
  volume    =  164,
  number    =  19,
  month     =  may,
  year      =  2026,
  language  = "en"
}

@ARTICLE{Taylor2024-ji,
  title     = "On the topological phase around conical intersections with
               Tamm-dancoff linear-response time-dependent density functional
               theory",
  author    = "Taylor, Jack T and Tozer, David J and Curchod, Basile F E",
  journal   = "J. Phys. Chem. A",
  publisher = "American Chemical Society (ACS)",
  volume    =  128,
  number    =  27,
  pages     = "5314--5320",
  month     =  jul,
  year      =  2024,
  copyright = "https://creativecommons.org/licenses/by/4.0/",
  language  = "en"
}

@ARTICLE{Taylor2023-tg,
  title    = "On the description of conical intersections between excited
              electronic states with {LR-TDDFT} and {ADC(2})",
  author   = "Taylor, Jack T and Tozer, David J and Curchod, Basile F E",
  journal  = "J. Chem. Phys.",
  volume   =  159,
  number   =  21,
  pages    = "214115",
  month    =  dec,
  year     =  2023,
  language = "en"
}

@ARTICLE{Wierichs2022-ta,
  title     = "General parameter-shift rules for quantum gradients",
  author    = "Wierichs, David and Izaac, Josh and Wang, Cody and Lin, Cedric
               Yen-Yu",
  journal   = "Quantum",
  publisher = "Verein zur Forderung des Open Access Publizierens in den
               Quantenwissenschaften",
  volume    =  6,
  number    =  677,
  pages     = "677",
  month     =  mar,
  year      =  2022,
  language  = "en"
}

@article{Mustroph2015,
author = {Mustroph, Heinz and Ernst, Steffen and Senns, Bianca and Towns, Andrew D.},
title = {Molecular electronic spectroscopy: from often neglected fundamental principles to limitations of state-of-the-art computational methods},
journal = {Coloration Technology},
volume = {131},
pages = {9-26},
doi = {https://doi.org/10.1111/cote.12120},
year = {2015}
}

@article{Pham2021,
author = {Pham, Thanh Chung and Nguyen, Van-Nghia and Choi, Yeonghwan and Lee, Songyi and Yoon, Juyoung},
title = {Recent Strategies to Develop Innovative Photosensitizers for Enhanced Photodynamic Therapy},
journal = {Chem. Rev.},
volume = {121},
pages = {13454-13619},
year = {2021},
doi = {10.1021/acs.chemrev.1c00381}
}

@article{Cardone2022,
  author    = {Cardone, Antonio and Capodilupo, Agostina Lina},
  title     = {Functional organic materials for photovoltaics},
  journal   = {Materials},
  year      = {2022},
  volume    = {15},
  pages     = {6333},
  doi       = {10.3390/ma15186333}
}

@article{Mohamadpour2024,
  author    = {Mohamadpour, Farzaneh and Amani, Ali Mohammad},
  title     = {Photocatalytic systems: reactions, mechanism, and applications},
  journal   = {RSC Advances},
  year      = {2024},
  volume    = {14},
  pages     = {20609--20645},
  doi       = {10.1039/D4RA03259D}
}

@article{Hamblin2004,
  author    = {Hamblin, Michael R. and Hasan, Tayyaba},
  title     = {Photodynamic therapy: a new antimicrobial approach},
  journal   = {Photochemical \& Photobiological Sciences},
  year      = {2004},
  volume    = {3},
  pages     = {436--450},
  doi       = {10.1039/B311900A}
}

@article{TAO2025102142,
title = {Engineering emissive excited states in organic electroluminescent materials},
journal = {Matter},
volume = {8},
pages = {102142},
year = {2025},
doi = {https://doi.org/10.1016/j.matt.2025.102142},
author = {Peng Tao and Jibiao Jin and Xiaokang Zheng and Yong-Jin Pu and Wai-Yeung Wong}
}

@article{helgaker2012,
	title = {Recent {Advances} in {Wave} {Function}-{Based} {Methods} of {Molecular}-{Property} {Calculations}},
	volume = {112},
	doi = {10.1021/cr2002239},
	journal = {Chem. Rev.},
	author = {Helgaker, Trygve and Coriani, Sonia and Jørgensen, Poul and Kristensen, Kasper and Olsen, Jeppe and Ruud, Kenneth},
	year = {2012},
	pages = {543--631},
}

@article{Saade2024,
  author    = {Saade, Sandra and Burton, Hugh G. A.},
  title     = {Excited State-Specific CASSCF Theory for the Torsion of Ethylene},
  journal   = {J. Chem. Theory Comput.},
  year      = {2024},
  volume    = {20},
  number    = {12},
  pages     = {5105--5114},
  doi       = {10.1021/acs.jctc.4c00212}
}

@article{Kossoski2023,
  author    = {Kossoski, Fábrys and Loos, Pierre-François},
  title     = {State-Specific Configuration Interaction for Excited States},
  journal   = {J. Chem. Theory Comput.},
  year      = {2023},
  volume    = {19},
  pages     = {2258--2269},
  doi       = {10.1021/acs.jctc.3c00057}
}

@article{Damour2024,
  author = {Damour, Yann and Scemama, Anthony and Jacquemin, Denis and Kossoski, F{\'a}bio and Loos, Pierre-Fran{\c{c}}ois},
  title = {State-Specific Coupled-Cluster Methods for Excited States},
  journal = {J. Chem. Theory Comput.},
  year = {2024},
  volume    = {20},
  pages     = {4129--4145},
  doi = {10.1021/acs.jctc.4c00034}
}

@article{Ivanov2006,
  author = {Ivanov, Valery V. and Adamowicz, Ludwik and Lyakh, Dmitry I.},
  title = {Excited states in the multireference state-specific coupled-cluster theory with the complete active space reference},
  journal = {J. Chem. Phys.},
  volume = {124},
  number = {18},
  pages = {184302},
  year = {2006},
  doi = {10.1063/1.2190221}
}

@ARTICLE{Anselmetti2021-ra,
  title     = "Local, expressive, quantum-number-preserving {VQE} ans{\"a}tze
               for fermionic systems",
  author    = "Anselmetti, Gian-Luca R and Wierichs, David and Gogolin,
               Christian and Parrish, Robert M",
  journal   = "New J. Phys.",
  publisher = "IOP Publishing",
  volume    =  23,
  number    =  11,
  pages     = "113010",
  month     =  nov,
  year      =  2021,
  copyright = "https://creativecommons.org/licenses/by/4.0/"
}

@BOOK{Barone2011-eo,
  title     = "Computational strategies for spectroscopy",
  editor    = "Barone, Vincenzo",
  publisher = "John Wiley \& Sons",
  month     =  nov,
  year      =  2011,
  address   = "Nashville, TN",
  language  = "en"
}

@ARTICLE{Taube2006-dv,
  title     = "New perspectives on unitary coupled‐cluster theory",
  author    = "Taube, Andrew G and Bartlett, Rodney J",
  journal   = "Int. J. Quantum Chem.",
  publisher = "Wiley",
  volume    =  106,
  number    =  15,
  pages     = "3393--3401",
  month     =  jan,
  year      =  2006,
  copyright = "http://onlinelibrary.wiley.com/termsAndConditions\#vor",
  language  = "en"
}

@ARTICLE{Kandala2017-nj,
  title     = "Hardware-efficient variational quantum eigensolver for small
               molecules and quantum magnets",
  author    = "Kandala, Abhinav and Mezzacapo, Antonio and Temme, Kristan and
               Takita, Maika and Brink, Markus and Chow, Jerry M and Gambetta,
               Jay M",
  journal   = "Nature",
  publisher = "Springer Science and Business Media LLC",
  volume    =  549,
  number    =  7671,
  pages     = "242--246",
  month     =  sep,
  year      =  2017,
  language  = "en"
}

@ARTICLE{Olsen1985-pg,
  title     = "Linear and nonlinear response functions for an exact state and
               for an {MCSCF} state",
  author    = "Olsen, Jeppe and Jørgensen, Poul",
  journal   = "J. Chem. Phys.",
  publisher = "AIP Publishing",
  volume    =  82,
  number    =  7,
  pages     = "3235--3264",
  month     =  apr,
  year      =  1985,
  language  = "en"
}

@ARTICLE{Christiansen1998-mb,
  title     = "Response functions from Fourier component variational
               perturbation theory applied to a time-averaged quasienergy",
  author    = "Christiansen, Ove and Jørgensen, Poul and Hättig, Christof",
  journal   = "Int. J. Quantum Chem.",
  publisher = "Wiley",
  volume    =  68,
  number    =  1,
  pages     = "1--52",
  year      =  1998,
  language  = "en"
}

@ARTICLE{Pawlowski2015-vd,
  title     = "Molecular response properties from a Hermitian eigenvalue
               equation for a time-periodic Hamiltonian",
  author    = "Paw{\l}owski, Filip and Olsen, Jeppe and J{\o}rgensen, Poul",
  journal   = "J. Chem. Phys.",
  publisher = "AIP Publishing",
  volume    =  142,
  number    =  11,
  pages     = "114109",
  month     =  mar,
  year      =  2015,
  language  = "en"
}

@ARTICLE{Rossi2026-sd,
  title         = "Resource-efficient energy-based operator selection in
                   fermionic {ADAPT-VQE} via exact Hamiltonian transformation",
  author        = "Rossi, Emanuele and Kjellgren, Erik Rosendahl and Izmaylov,
                   Artur F and Sauer, Stephan P A and Ziems, Karl Michael and
                   Coriani, Sonia",
  month         =  jun,
  year          =  2026,
  copyright     = "http://creativecommons.org/licenses/by-sa/4.0/",
  archivePrefix = "arXiv",
  primaryClass  = "quant-ph",
  eprint        = "2606.04786",
  journal  = "arXiv",
}

@ARTICLE{Ollitrault2020-jd,
  title     = "Quantum equation of motion for computing molecular excitation
               energies on a noisy quantum processor",
  author    = "Ollitrault, Pauline J and Kandala, Abhinav and Chen, Chun-Fu and
               Barkoutsos, Panagiotis Kl and Mezzacapo, Antonio and Pistoia,
               Marco and Sheldon, Sarah and Woerner, Stefan and Gambetta, Jay M
               and Tavernelli, Ivano",
  journal   = "Phys. Rev. Res.",
  publisher = "American Physical Society (APS)",
  volume    =  2,
  number    =  4,
  month     =  oct,
  year      =  2020,
  copyright = "https://creativecommons.org/licenses/by/4.0/",
  language  = "en"
}

@ARTICLE{Kumar2023-nf,
  title    = "Quantum simulation of molecular response properties in the {NISQ}
              era",
  author   = "Kumar, Ashutosh and Asthana, Ayush and Abraham, Vibin and
              Crawford, T Daniel and Mayhall, Nicholas J and Zhang, Yu and
              Cincio, Lukasz and Tretiak, Sergei and Dub, Pavel A",
  journal  = "J. Chem. Theory Comput.",
  volume   =  19,
  number   =  24,
  pages    = "9136--9150",
  month    =  dec,
  year     =  2023,
  language = "en"
}

@ARTICLE{Asthana2023-eu,
  title     = "Quantum self-consistent equation-of-motion method for computing
               molecular excitation energies, ionization potentials, and
               electron affinities on a quantum computer",
  author    = "Asthana, Ayush and Kumar, Ashutosh and Abraham, Vibin and
               Grimsley, Harper and Zhang, Yu and Cincio, Lukasz and Tretiak,
               Sergei and Dub, Pavel A and Economou, Sophia E and Barnes, Edwin
               and Mayhall, Nicholas J",
  journal   = "Chem. Sci.",
  publisher = "Royal Society of Chemistry (RSC)",
  volume    =  14,
  number    =  9,
  pages     = "2405--2418",
  month     =  mar,
  year      =  2023,
  copyright = "http://creativecommons.org/licenses/by/3.0/",
  language  = "en"
}

\end{document}